\documentclass[a4paper]{article}
\usepackage{epsfig,amsmath,amsfonts,amssymb,setspace,multirow,textcomp}
\usepackage[T1]{fontenc}
\usepackage{hyperref}
\makeatletter
\renewcommand{\@oddhead}{\textit{Advances in Astronomy and Space Physics} \hfil}
\renewcommand{\@evenfoot}{\hfil \thepage \hfil}
\renewcommand{\@oddfoot}{\hfil \thepage \hfil}
\makeatother

\renewenvironment{thebibliography}[1]{\begin{oldthebibliography}{#1}\setlength{\parskip}{0ex}\setlength{\itemsep}{0ex}}{\end{oldthebibliography}}

\usepackage[dvipsnames]{xcolor} 

\begin{document}
\fontsize{11}{11}\selectfont 
\title{VarStar Detect, a Python library dedicated to the semi-automatic detection of stellar variability}
\author{\textsl{P.\,G.~Jorge$^{1}$, C.\,A.~Nicolás$^{2}$,  C.\,B.~Andrés$^{3}$}}
\date{\vspace*{-6ex}}
\maketitle
\begin{center} {\small $^{1}$Department of Physics and Astronomy, University College London, Gower Street, WC1E 6BT London, United Kingdom \\
$^{2}$Facultad de Ciencias, University of Oviedo, C. Federico García Lorca, 18, 33007 Oviedo, Spain\\
$^{3}$Escuela de Ingeniería Informática, University of Oviedo, Calle Valdés Salas, 11, 33007 Oviedo, Spain\\
{\tt jorper.gonzalez@gmail.com , nicocarrizosaarias@gmail.com, andrescabla@gmail.com}}
\end{center}

\begin{abstract}
\texttt{VarStar Detect} is a Python package available on PyPI optimized for the detection of variability inside photometric measurements. Based off of the Least Squares method of regression, \texttt{VarStar Detect} calculates the amplitude of a Fourier Polynomial fit of the data as a measure of variability to assess if the star is indeed variable. This work shows the mathematical background of the package and an analysis of the code's functionality on TESS Sector 1 Data. \\[1ex]
\end{abstract}

\section*{\sc introduction}
\indent \indent At the time of writing, over 2 000 000 variable stars are catalogued in the Variable Star Index (VSX), currently run
by the American Association of Variable Stars Observers \footnote{Obtained from \url{https://www.aavso.org/vsx/index.php}}. Only a handful of these objects have been studied in full detail, with measurements of their physical parameters such as their mass, radii or temperature. Further, these variables are normally studied only once at the time of submission to the catalog \cite{Tkachenko2016}. These studies are usually performed with photometry since spectroscopic studies are more costly than those done with CCDs. Consequently, these studies adopt phenomenological approaches rather than physical ones. These analysis, required by the catalog previous to submission, classify stars by parameters such as its period $P$, initial epoch $T_0$ and magnitude range $m_{\text{max}} - m_{\text{min}}$. Variability type classification is performed manually by visual inspection of the light curve or phase plot, where the variability cycles are displayed with higher resolution.

The detection of variable star systems is important in astrophysics since it aids areas across the field: Cepheid variables are used as standard candles to calculate distances \cite{Majaess2009}. Further, detection of eclipsing binaries is essential to exoplanet hunting research due to its regular confusion with exoplanet transits by software \cite{Lissauer2014}.

In addition, the  amateur community can easily contribute to the detection of variable stars since said community obtains many CCD images for a variety of purposes (astrophotography, comet tracking, etc.). Since, aperture selection can be tedious work, many times these images are not inspected photometrically by the astronomer, specially if he/she is not interested in variable star detection. Furthermore, the majority of times, the astronomer will not encounter unknown variability in the brightness of the stars. This time consuming work therefore serves no purpose to the astronomer.

In this work we present \texttt{VarStar Detect}, a Python library dedicated to the semi-automatic detection of stellar variability. This library is thought to be the basis code of the \texttt{VarStar Detect} program (with implemented GUI) still to be written. This program will perform aperture photometry of all the targets in the FOV automatically and present the possible stellar variable candidates for later visual inspection by the astronomer. This way, the effort and time consumed in mining of variability in the astronomer's images will be significantly reduced, encouraging astronomers to look for variability in their data for submission to VSX. This paper presents the functionality and assessment of the \texttt{VarStar Detect} package available on PyPI\footnote{To access the package see \url{https://pypi.org/project/varstardetect/} for installation instructions or type \texttt{pip install varstardetect} in your command line. Full documentation of available functions and tutorials are available on the VarStar Detect github repository: \url{https://github.com/VarStarDetect/varstardetect}} for installation.

In the first part of this paper, a mathematical overview of the \texttt{amplitude\_test} function design is presented. The current version of \texttt{VarStar Detect} is optimized for the mining of TESS data.  In the latter part, a short investigation on sector 1 TESS light curves with \texttt{VarStar Detect} is shown to determine its performance.

\section*{\sc \texttt{VarStar Detect} design and mathematical background}

\indent \indent \texttt{VarStar Detect} is designed to detect variability in photometric light curves. To do that, it must accomplish four tasks:

\begin{enumerate}
    \item Fit a trigonometric polynomial of degree s to the light curve (using the weighted least squares method) and calculate its corresponding reduced $\chi^2$ parameter ($\chi^2_r$).
    \item Choose the trigonometric polynomial fit which $\chi^2_r\rightarrow 1$.
    \item Calculate the amplitude of the polynomial fit.
    \item Determine if star is variable.
\end{enumerate}

\indent \indent In the case we are dealing with, we are looking to identify a functional relationship between the flux intensity with respect to the time variable. This relationship should illustrate the behaviour of the obtained data and, from it, the characteristics of the star in question.

\subsection*{\sc data processing flow}
\indent \indent The primary input for \texttt{VarStar Detect} are lists of time, flux and flux uncertainty measurements\footnote{Which are properly filtered by the program to trim errors and non-existant data.}. Time is supported in any unit, although Heliocentric Julian Date (HJD) is recommended for submission to VSX. Brightness measurements (flux) are recommended to be inputted in magnitudes, which can be easily obtained with:
\begin{equation}
    m = \overline{m} - 2.5\log{\left(\frac{F}{\overline{F}}\right)}
\end{equation}
where m is the magnitude, F is the flux, $\overline{m}$ is the average magnitude of the star easily obtained from VizieR\footnote{\href{https://vizier.u-strasbg.fr}{https://vizier.u-strasbg.fr}} and $\overline{F}$ is the average flux value of the whole data set.

Given the physical context of the problem, it is known that the relationship sought must be sinusoidal \cite{Andronov2012}. Therefore, we will contemplate the following parametrization of Fourier's Development as the approximating function:

\begin{equation}
    P_s(t) = C_1 + \sum_{i = 1}^s \left( C_{i,1} \cdot \cos(i\* \omega \* (t - T_0)) + C_{i,2} \cdot \sin(i\* \omega \* (t - T_0))  \right)
\end{equation}
where $\omega$ is the angular frequency of the function (given by the Lomb-Scargle periodogram \cite{VanderPlas2018}) and $T_0$ the initial epoch, characteristics of the star. In summary, it consists on a $s$-degree trigonometric polynomial with coefficients $C_1,C_{1,1},\dots,C_{s,2}$ to be determined through regression.

Due to the sheer amount of data and knowledge of the uncertainty on the flux's measurements, the regressive method of weighted least squares shall be applied. Not only because of its computational ease, but also because using interpolating techniques would lead to Runge's phenomenon, providing a fake tendency of the data (its variability should be taken into account).

The weighted least squares method consists on a generalization of the widely-known least squares method. By taking into account the uncertainty of each data point, it provides a more accurate approximation function which, differently than the original method, minimizes the $\chi^2$ parameter:
\begin{equation}
    \chi^2 = \sum_{j = 1}^n \left(  \frac{y_j - P_s(x_j)}{\Delta y_j} \right)^2
\end{equation}
where $n$ is the number of data points and $\Delta y_j$ the uncertainty of the $y_j$ measurement.

Considering the minimization of the derivative of $\chi^2$ with respect to each parameter, the following lineal system is obtained, given by the normal equations:

\begin{equation}
    (X^T W X)\cdot \beta = (X^T W Y)
\end{equation}
where $\beta =$\scalebox{0.75}{$ 
\begin{pmatrix} C_1 \\ C_{1,1} \\ \vdots \\ C_{s,2} \end{pmatrix}$} is the coefficients matrix, 
$X =$ \scalebox{0.75}{$
\begin{pmatrix}
1 & \cos(\omega(t_1 - T_0)) & \sin(\omega(t_1 - T_0)) & \hdots & \sin(s\omega(t_1 - T_0))\\
1 & \cos(\omega(t_2 - T_0)) & \sin(\omega(t_2 - T_0)) & \hdots & \sin(s\omega(t_2 - T_0))\\
\vdots & \vdots&\vdots &\ddots &\vdots \\
1 & \cos(\omega(t_n - T_0)) & \sin(\omega(t_n - T_0)) & \hdots & \sin(s\omega(t_n - T_0))
\end{pmatrix}$}
 the function matrix,  
$W =$\scalebox{0.75}{$
\begin{pmatrix}
{1}/{\Delta y_1 ^2} & 0 & \hdots & 0 \\
0 &{1}/{\Delta y_2 ^2} & \hdots & 0 \\
\vdots & \vdots & \ddots & \vdots\\
0 & 0& \hdots & {1}/{\Delta y_n ^2}
\end{pmatrix}$}
  the weights matrix and $Y =$\scalebox{0.75}{$
\begin{pmatrix}
y_1  \\
y_2 \\
\vdots \\
y_n
\end{pmatrix}$}
is the flux matrix.

The solution to this system provides the optimal coefficients ($\beta$) which minimize $\chi^2$.
 
\texttt{VarStar Detect} then fits the data to several polynomials of degree $s\in [1,30]$ and calculates its respective $\chi^2_r$ parameter (current version of \texttt{VarStar Detect} is optimized for TESS, which observes stars for 30 days, the maximum detectable period is therefore 30 days).

Although the principle of maximum likelihood suggests  $\chi^2$ minimization, to determine the degree  $s$ of the polynomial, the $\chi^2_r$ parameter will be considered:

 \begin{equation}
     \chi^2_r = \frac{\sum_{j = 1}^n \left(  \frac{y_j - P_s(x_j)}{\Delta y_j} \right)^2}{n - 2\*s -1}
 \end{equation}
 \indent \indent This parameter takes into account the degrees of freedom of the fit ($n - 2\*s -1$, since the $s$-degree polynomial has $2\*s +1$ terms).
 Instead of minimising $\chi^2_r$ , we will pursue $\chi^2_r \to 1$ , to avoid the function from over-fitting. 
 
\texttt{VarStar Detect} then obtains the amplitude of the polynomial by subtracting the lowest fitted value from the highest:

\begin{equation}
    A = \texttt{max}(P_s(x_j)) - \texttt{min}(P_s(x_j))
\end{equation}

To obtain the uncertainty of the amplitude, knowledge of the uncertainty of the fitted values is necessary and therefore the uncertainty of the fitting parameters are needed. In a generic regression, to identify the parameter's uncertainty, the covariance matrix ($cov$) needs to be constructed, gathering the covariance of each pair of parameters.

In our case, we will only be interested in the diagonal elements of $cov$, and following the method described  \footnote{Obtained from \url{https://www.gnu.org/software/gsl/doc/html/lls.html\#c.gsl_multifit_wlinear}}, we can calculate

\begin{equation}
    cov = (X^t W X)^{-1}
\end{equation}

easily obtained with the \textit{numpy} package \footnote{\href{https://numpy.org}{https://numpy.org}}.

\indent\indent  Following the general theory described in \cite{MesandUncer}, to calculate the uncertainty of the fitted function, it is enough to consider that the  given function is a vectorial function in which, aside from the time as an independent variable, the parameters are also considered as variables:
$$ 
P_s(t) \Rightarrow P_s(t, C_1, C_{1,1}, \dots, C_{s,2})
$$

Then, since $\Delta t = 0$ (it's not a measurement), we will consider:

\begin{equation}
    \Delta P_s(z) = \sqrt{\left( \frac{\partial P_s}{\partial C_1}(z)\cdot \Delta C_1 \right)^2 +
    \left( \frac{\partial P_s}{\partial C_{1,1}}(z)\cdot \Delta C_{1,1} \right)^2 + 
    \dots +
    \left( \frac{\partial P_s}{\partial C_{s,2}}(z)\cdot \Delta C_{s,2} \right)^2
    }
\end{equation}

For each parameter, the derivative is just a single element of the given vectorial function evaluated in $z$ and, therefore, the calculus of this uncertainty can be easily obtained computationally with a chained loop.

\indent\indent With a similar expression to the above, the amplitude uncertainty can be calculated taking into account the accumulated error on a subtraction.

 \begin{equation}
     \Delta A = \sqrt{\left(\Delta\texttt{max}(P_s(x_j))\right) ^ 2 + \left(\Delta\texttt{min}(P_s(x_j))\right) ^ 2 }
 \end{equation}

Knowing these values, \texttt{VarStar Detect} proceeds with the amplitude variability test.

\subsection*{\sc amplitude test for stellar variability detection}
\indent \indent The purpose of the amplitude test is to decide if the star in question is a potential variable candidate that deserves to be visually inspected. In essence, it nominates as potential variables all stars which have an amplitude greater or equal than the threshold amplitude. The question comes when selecting the threshold amplitude. This will depend on the equipment used and quality of the night sky. It measures the smallest amplitude the equipment can detect before measuring just noise. This work does not go into explaining the physical meaning of this threshold. We encourage further investigations to do so. In the testing of the code, an empirical approach was used to determine the value of threshold for the data. 

\section*{\sc test on tess}
\subsection*{\sc transiting exoplanet survey satellite (tess)}
\indent \indent TESS is a space telescope launched by NASA on the 18$^{\text{th}}$ of April 2018 aboard a SpaceX Falcon 9 rocket \cite{ricker2014}. Its scientific objective was similar to those of its predecessor (Kepler): to find extrasolar planets and to calculate their mass and atmospheric compositions.

To achieve the previously outlined objective, aboard the telescope lies four identical CCDs that together monitor sectors of 24º$\times$ 90º of the sky. Each sector is observed for a total of 27 days, which consist of two full orbits around Earth. During this time, the telescope produces photometry with a 2 minute cadence. 

Furthermore, TESS data is available at different processing levels \cite{Fausnaugh2018}. Simple Aperture Photometry flux (SAP) is the performed photometry without removal of systematic variations by removal of common variability signals to all stars. Data with this level of correction is the Pre-search Data Conditioning Simple Aperture Photometry flux (PDCSAP). In this study, PDCSAP flux will be used for data mining and SAP will be used when determining if phenomena has been erroneously introduced by the photometry extraction process or astrophysical event. 

In this study we have analysed the first 500 light curves of the first sector observed by TESS to determine the efficiency of \texttt{VarStar Detect}. 

\subsection*{\sc Results}
\indent \indent Firstly, 3$\sigma$ ourliers were looked for in the data. All of these values were removed from the LC files downloaded using the \texttt{lightkurve} package \footnote{\href{https://docs.lightkurve.org}{https://docs.lightkurve.org}}. PDCSAP flux was used.

Following the outlier extraction process, the evaluation test function was applied to obtain possible variable star candidates. For TESS data, a threshold of 20 e$^-$s$^{-1}$ was selected. This threshold was chosen after visual inspection of variable star light curves discovered through TESS. Random lightcurves (selected with the \texttt{random} \footnote{\url{https://docs.python.org/3/library/random.html\#module-random}} python function) were inspected. Highest noise values were about 20 e$^-$s$^{-1}$.

500 light curves were analysed with the amplitude test function. Out of the 500 stars surveyed, a total of 169 were nominated as variables. After visual inspection of the variable candidate lightcurves, a total of 163 present periodic variability in brightness over time. This results in a efficiency of 96.45 $\%$ in the program's functionality.

This test run also detected exoplanet transit WASP-126 c which had been previously discovered \cite{Pollacco2006}.

All these detected variable stars will be submitted to VSX after checking on VizieR if they had been previously discovered by other teams.

\section*{\sc conclusion and discussion}
\indent \indent In this work, we have presented \texttt{VarStar Detect}: a python dedicated library to semi-automatic detection of stellar variability. The mathematical background behind the python package was described, the functionality of the amplitude test was introduced and the program as a whole was tested on Sector 1 TESS PDCSAP data, downloaded with the data download function designed and available in the python repository. The program was proved to be 96.45 $\%$ efficient, detecting a total of 163 variable stars inside the imported sector 1 database. Following up this analysis, catalogation of these objects is being performed.

\texttt{VarStar Detect} is a program which focuses on amateur variable star detection. Since submission to the catalog requires visual inspection of the data, the program is dedicated to candidate extraction from an imported database. The current version of \texttt{VarStar Detect} is optimised for stellar variability detection using TESS. Future versions of \texttt{VarStar Detect} will include a function of aperture photometry not yet tested at the moment which will allow import of databases in the form of .FIT images.

Furthermore, as previously mentioned, this investigation does not present a quantitative method for determining the value for the threshold of the amplitude test in an objective manner and a further study is encouraged.

In sum, \texttt{VarStar Detect} is a simple python package that facilitates variable star detection at high precision. 

\section*{\sc acknowledgement}
\indent \indent The development of this software comes from the voluntary initiative of the three undergraduate authors. It did not count with any academic supervision except for the advice of the following academics we want to acknowledge: Dr. Isabel Llorente García (UCL), Dr. Pablo Pérez Riera (University of Oviedo) and Dr. Carlos Enrique Carleos Artime (University of Oviedo). Also, thank you Ellen for your support in writing this paper.

\end{document}